\documentclass[runningheads]{llncs}
\usepackage[T1]{fontenc}
\usepackage{cite}
\usepackage{amssymb,amsfonts}
\usepackage{algorithmic}
\usepackage{graphicx}
\usepackage{textcomp}
\usepackage{xcolor}
\usepackage{romannum}
\usepackage{algorithmic}
\usepackage[ruled,vlined,linesnumbered]{algorithm2e}
\usepackage{float}
\usepackage{subfigure}
\usepackage{orcidlink}
\usepackage{xspace}
\usepackage[colorinlistoftodos]{todonotes}
\usepackage{amsmath}
\usepackage{booktabs}
\usepackage{multirow}
\newcommand{\MILAAP}{MiLAAP\xspace}

\begin{document}
%
\title{\MILAAP: Mobile Link Allocation via Attention-based Prediction}

\titlerunning{Mobile Link Allocation via Attention-based Prediction}
%
\author{Anonymous Author(s)}
\author{Yung-Fu Chen\orcidlink{0009-0002-5331-409X} \and
Anish Arora\orcidlink{0000-0003-2152-0462}}

\authorrunning{Y. Chen et al.}
%
\institute{The Ohio State University, Columbus, OH 43210, USA \\ \email{\{chen.6655, arora.9\}@osu.edu}
}
\maketitle              
\vspace{-6mm}
\begin{abstract}
To maintain throughput efficiency in dynamic wireless networks, channel hopping communication systems must  adapt continuously as interference evolves and nodes move. While optimal scheduling requires up-to-date channel occupancy information to select non-overlapping channels, sharing this network state among nodes introduces significant communication overhead. This overhead severely degrades the performance of already capacity-limited networks, especially as network size or node mobility scales. To address this, we propose \MILAAP, a novel attention-based framework that eschews explicit state sharing in favor of localized, learning-based channel occupancy prediction. By employing a self-attention mechanism, \MILAAP enables each node to independently capture temporo-spectral channel hopping patterns within its interference region and predict local channel occupancy states. Because it relies entirely on passively observed local channel activities, the framework introduces zero communication overhead. Furthermore, to handle node mobility, \MILAAP integrates a multi-head self-attention mechanism that captures temporo-spatial dependencies, allowing nodes to predict the motion trajectories of interfering neighbors. Detecting when nodes enter or exit an interference region further refines the accuracy of these channel occupancy predictions. \MILAAP is inherently stabilizing in that if the channel hopping sequence of one or more neighbors of a node changes, the node recaptures the new channel hopping pattern in $2L$ time slots (where $L$ is the channel hopping sequence period). Our evaluations consider both terrestrial wireless meshes and satellite meshes. We show that \MILAAP achieves $\sim$100\% channel occupancy state prediction accuracy across diverse mobility models in both terrestrial wireless meshes and satellite meshes. Moreover, \MILAAP demonstrates zero-shot generalizability across various channel hopping sequence periods and node mobility patterns.

\vspace{-2mm}
\keywords{machine learning \and channel hopping \and mobile networks \and satellite meshes \and self-attention \and sequence to sequence model \and mobility detection \and interference avoidance \and stabilization.}
\end{abstract}
\section{Introduction}
\vspace{-2mm}
Channel hopping (also known as frequency hopping) has been widely adopted in wireless communication systems to mitigate interference and resist active jamming or passive eavesdropping attacks. It combines a slotted time schedule for multi-channel medium access with a channel hopping mechanism to effectively avoid collisions in concurrent transmissions. As demand for wireless network applications increases, channel hopping communication systems must evolve to handle network dynamics. 

The evolution of channel hopping communication systems is particularly important for low-rate and low-power communication systems, such as the Industrial Internet of Things (IIoT) \cite{mohamadi2020industrial} and Edge Internet of Things \cite{rekha2025tsch}, where channel hopping has gained significant adoption but remains inadequately equipped to manage dynamic interference and mobility. 
In these environments, Time Slotted Channel Hopping (TSCH) is frequently employed. While the IEEE 802.15.4 standard allows implementers to choose between centralized and distributed scheduling \cite{mohamadi2018scheduling}, TSCH networks typically rely on a globally fixed, long channel sequence ($\geq$ 256)---selected from a sufficient number of available channels---along with an offset schedule to avoid interference and resist jamming attacks \cite{cheng2020cracking}. This rigid reliance on a global sequence severely limits robustness to network dynamics, particularly because TSCH lacks built-in mechanisms to manage node mobility. Furthermore, the requirement for a large channel set undermines the feasibility of TSCH in resource-constrained networks. Although a few mobility-aware protocols have been proposed \cite{tavallaie2021design, al2015mobility, pettorali2024mobility}, adapting to such dynamics traditionally requires proactive spectrum sensing and continuous message exchanges among neighboring nodes to share up-to-date network states. Unfortunately, this frequent message passing introduces significant communication overhead, which degrades throughput efficiency—especially as network size or node mobility scales. Consequently, achieving efficient, mobility-aware scheduling without the crippling overhead of network state synchronization remains a critical and unresolved challenge.

Beyond terrestrial and low-power industrial frameworks, channel hopping has been adopted in long-range satellite communication systems. Satellite networks, such as Intelsat~\cite{Intelsat}, Inmarsat~\cite{Inmarsat}, and Eutelsat~\cite{fenech2016eutelsat}, utilize low Earth orbit (LEO) satellites to offer lower latency and higher bandwidth, alongside geostationary Earth orbit (GEO) satellites to provide broad coverage. Channel hopping is applied to communication links between LEO/GEO satellites and their ground stations and the inter-satellite links (ISLs) between LEO and GEO satellites \cite{aykin2018adaptive}. Since rapid changes in environmental conditions cause cross-channel interference to alter materially, dynamic channel hopping \cite{aykin2018adaptive} addresses the limitations of static configurations to significantly improve communication resilience, security, interference mitigation, and cost-effectiveness. However, such cognitive radio based dynamic schemes necessitate a centralized coordinator to manage frequency assignments using proactive out-of-band sensing and global link metrics. This reliance on explicit control messaging and centralized infrastructure introduces significant communication overhead, highlighting the critical need for localized, zero-overhead sequence scheduling.

\subsection{Limitations of Existing Predictive Models}
To eliminate this communication overhead, dynamic channel hopping can be framed as a localized state prediction problem. However, existing approaches fail to provide a comprehensive solution for mobile networks:  
\begin{enumerate}

\item {\em Traditional Time-Series}:~~~The Auto Regressive Integrated Moving Average (ARIMA) model fails when the period of the channel hopping sequence exceeds the fixed length of the look-back window. Similarly, Gaussian Process Regression (GPR) struggles as the discrete, repeating patterns of channel hopping sequences are poorly captured by continuous Gaussian process modeling.  

\item {\em Historical Predictors}:~~~Markov Chain Predictors and Pattern Repeaters (Autocorrelation) can achieve near 100\% accuracy in static networks. However, they suffer a drop in accuracy when applied to mobile networks because they inherently lack the mechanism to detect spatial dynamics, specifically, the location changes and trajectories of interfering neighboring nodes.  

\item {\em Machine Learning}:~~~Previous works have used long short-term memory (LSTM) \cite{li2020lstm}, convolutional neural network (CNN)\cite{lee2020detection}, and gated recurrent units (GRU)\cite{chung2014empirical} models to predict sequences. However, they have not adequately addressed the network dynamics of node mobility or external interference.

\end{enumerate}

\begin{figure}[t!]
    \centering
    \vspace{-4mm}
     \subfigure
     {
        \includegraphics[width=0.95\textwidth]{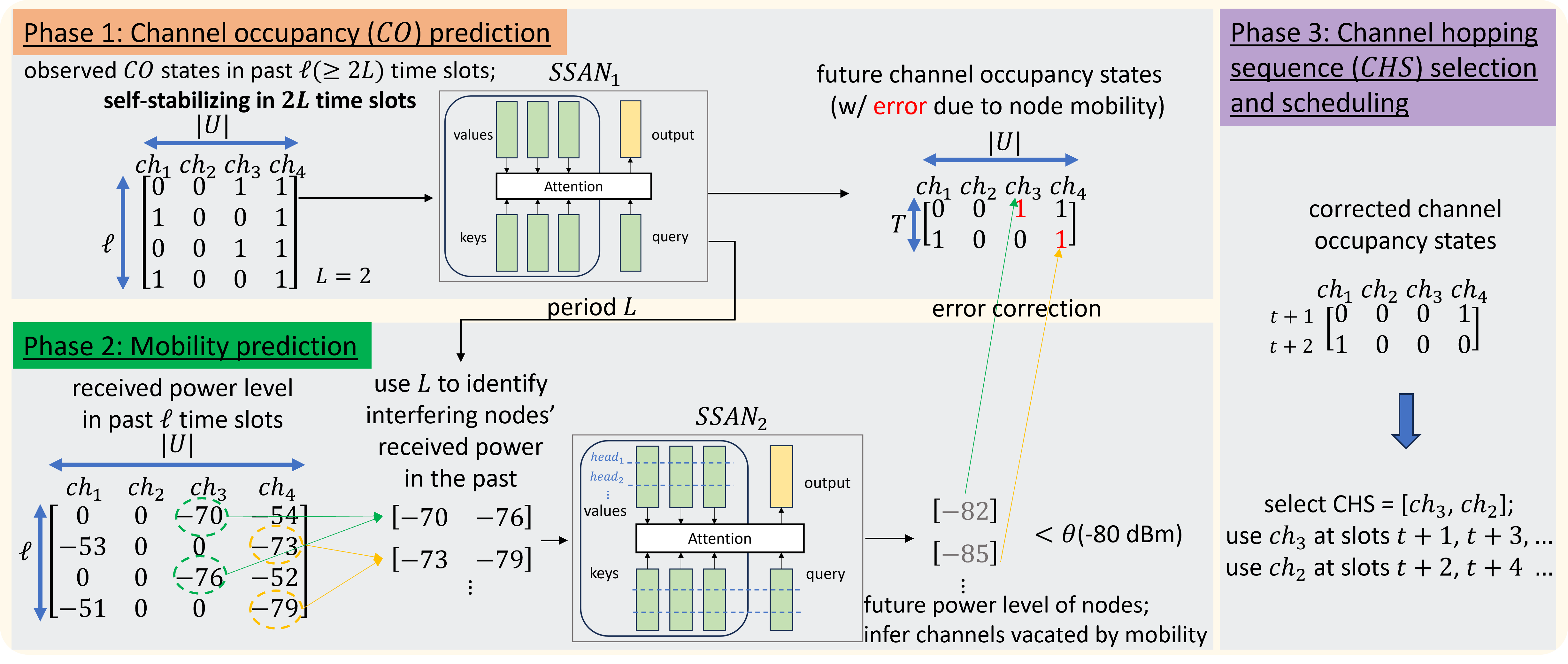}
    }
    \vspace{-4mm}
    \caption{Illustration of the procedure for mobile link allocation in \MILAAP: Prediction of occupancy of the $|U|$ channels by a sequence-to-sequence attention network (Phase 1) identifies unoccupied channels that are used to select the node's new local channel hopping sequence of period $L$ (Phase 3). As an intermediate step, prediction of received power level from other nodes by another sequence-to-sequence attention network (Phase 2) is used to detect entry or departure of nodes from the node's interference region and expedite the correction of predicted channel occupancy states.}
    \label{MILAAP}
\end{figure}

\subsection{The \MILAAP Approach}
To address these limitations, we formulate a purely local approach to predict channel occupancy (CO) states and allocate communication channels with minimal collision. We propose \MILAAP\footnote{"\MILAAP" means coming together in unison in Hindi.}, an attention-based sequence-to-sequence learning architecture that eschews explicit network state sharing. The key objective of \MILAAP is to rapidly capture the temporal, spectral, and spatial dependencies between mobile nodes using only historical data observed locally and passively. The channel occupancy state at a given node results from the superposition of the channel hopping sequences of all transmitters within its interference range. Therefore, predicting this state requires capturing complex temporo-spectral correlations. Furthermore, because this state dynamically changes as other nodes move in and out of the interference region, predicting node mobility is equally crucial and involves capturing temporo-spatial correlations. \MILAAP solves this dual prediction problem in two phases (as depicted in Fig.~\ref{MILAAP}): the first phase addresses channel occupancy prediction by capturing the regular period of the superposed channel hopping sequences, while the second phase addresses mobility detection by predicting the received power levels of neighboring nodes to infer whether they are vacating or entering the interference region.

The attention mechanism has proven highly effective at learning complex temporo-spectral and temporo-spatial dependencies by dynamically weighting the significance of past observations \cite{lee2019spectro, ma2019attention}. \MILAAP leverages this capability by adopting an attention-based sequence-to-sequence learning architecture \cite{cao2022spatio, yang2024multistep}. For the first phase, a self-attention mechanism is ideally suited to identify repeating temporo-spectral patterns hidden within the superimposed CO history; it can directly connect distant time slots to recognize periodicity without the strict look-back constraints of traditional time-series models. For the second phase, \MILAAP utilizes multi-head self-attention to simultaneously track the temporo-spatial trajectories of multiple interfering neighbors. By allowing different attention heads to focus on distinct representational spaces---such as the received power levels of different nodes across various channels---the architecture can reliably detect mobility-driven state changes and swiftly adapt its predictions.


\subsection{Contributions of the paper} 
We highlight four aspects of \MILAAP: 
    \begin{enumerate}
    \item 
    {\bf Purely Local, Zero-Overhead Prediction}: \MILAAP predicts channel occupancy states at each node using only locally and passively observed information. This yields higher throughput efficiency compared to MAC protocols that rely on proactive or reactive message-passing overhead.
    \item 
    {\bf Temporo-spectral Pattern Capture}: A single-layer sequence-to-sequence self-attention network accurately captures the superposition of local channel hopping sequences, achieving near-perfect accuracy in static environments and providing zero-shot generalizability across varying sequence periods.
    \item  
    {\bf Mobility Detection via Temporo-spatial Dependencies}: By utilizing a multi-head self-attention layer, \MILAAP accurately tracks node trajectories by predicting received power levels. This enables the framework to adapt its schedules to dynamic mobility patterns seamlessly, maintaining near 100\% prediction accuracy even in highly mobile networks. We also show that learning is feasible for different node mobility patterns. 
    \item  
    {\bf High Efficiency and Performance}: Designed for practical deployment in resource-constrained networks, our evaluations, conducted using short channel hopping sequences ($L=4$) selected from a limited number of channels ($|U|=8$), demonstrate that \MILAAP achieves $\sim \!\! 100\%$ accuracy for predicting channel occupancy within a node’s local neighborhood in terrestrial wireless meshes and satellite meshes. Compared to Autocorrelation and Markov Chain Predictors, \MILAAP maintains the same baseline computational complexity for extracting patterns and outputting future sequences\footnote{Furthermore, the attention-based matrix calculations can be highly parallelized using GPUs, reducing the computational complexity to a logarithmic scale.}.
    \item
    {\bf Self Stabilization}: \MILAAP is intrinsically self-stabilizing; if one or more neighbors alter their transmission sequences, a node will successfully adapt to the updated channel pattern in $2L$ time slots.
\end{enumerate}

\section{Background}
\vspace{-2mm}
\subsection{Time Slotted Channel Hopping (TSCH)}


Time Slotted Channel Hopping (TSCH) is a robust channel hopping MAC protocol formally incorporated into the IEEE 802.15.4e standard \cite{IEEE802.15.4}. By inheriting slotted time access from the baseline IEEE 802.15.4 standard, TSCH effectively minimizes internal interference and mitigates multipath fading. The communication structure relies on a synchronized slotframe, defined as a repeating collection of fixed-length time slots. Each individual slot provides a duration sufficient to accommodate both a data packet transmission and its corresponding acknowledgment between a pair of nodes within a reliable transmission radius.  To avoid communication collisions, TSCH maps links onto specific channel offsets using a globally shared, predefined channel hopping sequence. This structural rigidity is primarily designed to minimize the network overhead associated with channel synchronization. Mathematically, the communication channel assigned to a given link at an Absolute Slot Number ($ASN$) is governed by the following function:

\vspace{-4mm}
\begin{equation}
\begin{array}{l}
Channel(ASN) = CHS \left[(ASN + OF) \ mod \ L \right] ,
\label{CHS}
\end{array}
\vspace{-2mm}
\end{equation}
where $\text{OF}$ denotes the channel offset designated to separate potentially interfering links, and $L$ represents the period of the channel hopping sequence. While various centralized and distributed scheduling heuristics \cite{hermeto2017scheduling} have been proposed to exploit frequency and time diversity to improve end-to-end reliability , the fundamental reliance on a globally uniform sequence restricts TSCH's native capacity to dynamically adapt to structural network changes or uncoordinated node mobility.


\subsection{Local Channel Hopping Sequences} 
To overcome the flexibility constraints of globally synchronized sequences, localized channel hopping strategies have been introduced to handle volatile topologies and shifting traffic demands. When the sequence period $L$ is small, global sequences struggle to suppress concurrent localized interference. Specifically, if the number of active communication links within an individual interference region exceeds $L$, predictable and recurrent collisions become unavoidable. This drops the achievable capacity by a factor of at least $\frac{1}{L}$, a degradation far more severe than the statistical average observed when deploying randomized channel sequences.

Prior research has demonstrated \cite{chen2022qfmac} that when the number of available channels $|U|$ is greater than or equal to the count of active nodes $K$, which in turn is greater than or equal to the number of concurrent flows (i.e., $|U| \geq K \geq \#(\text{flows}) > 1$), scheduling via dedicated, per-link local channel hopping sequences yields throughput competitive with or better than both global and random allocation strategies. However, existing methodologies require active, explicit coordination and message exchange among neighboring nodes to assign these localized sequences. While this active coordination permits localized adaptation to interference dynamics, it introduces significant synchronization overhead. To date, the feasibility of entirely passive, localized sequence selection---devoid of control-plane message sharing---remains unaddressed in the literature, motivating the machine learning-driven approach proposed in this work.

\section{Problem Statement}
\vspace{-2mm}
The primary objective of mobile link allocation is to efficiently assign limited network resources---communication channels---to mobile nodes in a way that minimizes interference across all active links within each node's interference region. High efficiency requires that the overhead introduced by channel selection and scheduling remains minimal. To achieve this, we decouple the objective into two sequential tasks: (1) accurately and efficiently predicting the local channel occupancy ($CO$) state at each node, and (2) leveraging these predictions to autonomously select a local channel hopping sequence ($CHS$) that avoids collisions. Fundamentally, minimizing these collisions by assigning non-overlapping channels for transmission is an instance of the graph coloring problem \cite{palattella2012traffic}.

\subsection{System Model}
The communication system uses slotted-time access, multi-channel communication, and channel hopping based on a synchronized slotframe. Each slotframe is a collection of fixed-length slots that repeats over time; every slot is long enough to transmit a data packet and receive an acknowledgment via a link between a pair of nodes within a reliable transmission radius. Each link corresponds to two nodes; without loss of generality, we assume that each link is assigned to use one channel at a given time slot.

\subsubsection{Communication Model}
A local channel hopping sequence $CHS_i$ is associated with each node $i$. $CHS_i$ determines the communication channel used by node $i$ to transmit at time slot $t$:
\vspace{-4mm}
\begin{equation}
\begin{array}{l}
Channel_i(t) = CHS_i\left[t \ mod \ L\right],
\label{CHS2}
\end{array}
\vspace{-2mm}
\end{equation}
where $L$ is the period (or, equivalently, the length) of $CHS_i$. The channel hopping sequence can be different for different nodes. 

Let $K$ be the number of active nodes transmitting traffic within an interference region. To avoid interference, $L$ must be at least $K$.
For ease of presentation, let us assume that all nodes use a common, fixed $L$. In this case, the following thought-experiment provides intuition for the local predictability of the channel occupancy in the interference region of a node: Let all nodes transmit in each slot. Then the channel occupancy at node $i$ will be the superposition of the channel hopping sequences of all nodes in the interference region of $i$. This channel occupancy will have a period of $L$ and can hence be predicted by $i$ by observing the channel occupancy in $2L$ time slots.\footnote{The use of different $L$ in an interference region is likewise feasible: in the same thought experiment, the predictor can capture the period by observing time slots over at least twice the least common multiple of the different $L$ used by the nodes.}

To observe channel occupancy states in a time slot $t$, $CO_t(i)$, node $i$ samples the medium in the channels to determine whether or not some signal is present. (The samples could be based on typical radio mechanisms such as Clear Channel Assessment (CCA), Received Signal Strength Indicator (RSSI), Channel State Indicator (CSI), or Signal to Noise Ratio (SNR).) For simplicity, we assume that all channels in $U$ can be sampled in any given time slot.\footnote{In practice, the sampling can be done in batches of size $B_i, B_i\geq1$, collected every $U/B_i$ time slots, and the number of time slots to be observed for prediction can be scaled accordingly.}  


\subsubsection{Traffic Model}
We assume the network supports traffic flows with moderate arrival and departure rate. In other words, internal interference in static networks changes when a flow arrives or departs but, by assumption, the duration between changes is larger than the duration of channel occupancy prediction and channel hopping sequence update. 


\subsubsection{Mobility Model}
Each network node can move within the network area, controlling its velocity and direction independently and unbeknownst to other nodes. Similar to the traffic model, we assume the rate at which nodes enter or leave interference regions is slower than the time required for $CO$ prediction and $CHS$ updates.

\begin{figure}[hbt!]
    \centering
     \subfigure
     {
        \includegraphics[width=0.8\textwidth]{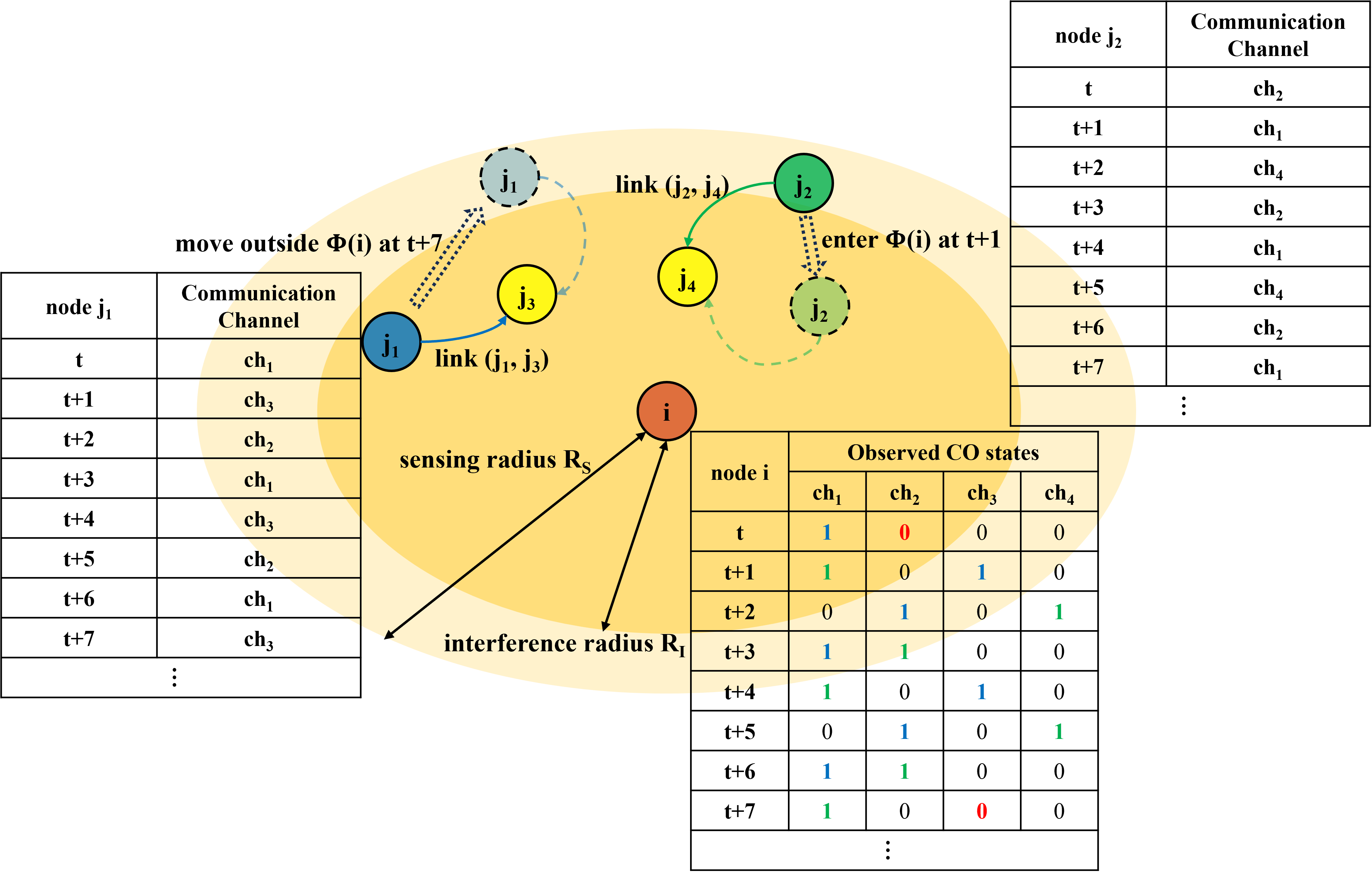}
    }
    \caption{Example of channel occupancy ($CO$) vectors observed by node $i$ in a mobile network with $L = 3$. The channel hopping sequences of nodes $j_1$ and $j_2$ are $CHS_{j_1}=ch_1;ch_3;ch_2$ and $CHS_{j_2}=ch2;ch1;ch4$. From time $t+1$, when both $j_1$ and $j_2$ are in $i$'s interference, the $CO$ vector repeats every $L$ time slots until $t+7$, when $j_1$ leaves the region. $i$ can detect the entry of $j_2$ into the region at $t+1$ from $CO_t$ and $CO_{t+1}$ and predict the leave of $j_1$ at $t+7$ from the received power trajectory in preceding slots.}
    \label{observed_channel_occupancy}
\end{figure}

\subsection{Predicting Channel Occupancy in Static and Mobile Networks}
\label{sec:PCO}
To maximize prediction accuracy, node $i$ learns both long-term temporo-spectral dependencies and dynamical temporo-spatial dependencies. For the former, node $i$ learns a model that predicts future channel occupancy states (i.e., the period of channel hopping sequences) associated with active nodes within $i$'s interference region. For the latter, $i$ tracks node mobility via the received power level from nodes $j$ in its interference region and accordingly predicts when nodes enter or leave its interference region.\footnote{Recall that the schedule for each node $i$ can be determined based on the prediction that $i$ makes for channel occupancy. 
The details of how the schedule of $i$ is selected given the prediction, including how and when the channel hopping schedule is updated and how scheduling efficiency is optimized, are outside the scope of our presentation. Here, we focus on defining the prediction sub-problems.} Fig.~\ref{observed_channel_occupancy} illustrates an example of how channel occupancy in the interference region of a node $i$ exhibits cyclic behavior---during time $t+1$ and $t+6$ where the set of nodes in the region remains unchanged. It also shows how the channel occupancy changes as nodes---$j_2$ at time $t+1$---move in and---$j_1$ at time $t+7$---move out of the interference region of node $i$. Predicting node mobility helps to infer when active nodes leave or enter the interference region of a node.

\subsubsection{\bf Channel Occupancy Prediction}
Let $ CO_t = \left[o_{ch}  | ch \! \in \! U\right]$ denote the vector of channel occupancy states observed by node $i$ at time slot $t$, where $o_{ch}\in \{0,1\}$ is 1 if channel $ch$ is occupied at $t$, 
and $U$ is the set of all channels. We formulate the learning problem for channel occupancy prediction as follows:
\begin{quote}
Given $\ell$ historical vectors of channel occupancy states, $CO_{t-\ell+1}, ..., CO_{t}$, observed by node $i$ within its interference region, a channel occupancy prediction model is learned to predict $T$ future vectors $CO_{t+1}, ..., \\ CO_{t+T}$. 
\end{quote}

\subsubsection{\bf Received Power Level Prediction for Mobility Detection}
To enable inference of the time slot at which an active node is entering or moving out of $i$'s interference region, we formulate a learning problem for received power level prediction as follows:

\begin{quote}
Given $\ell$ historical states of received power level from node $j \in \Phi(i)$, $s_{t-\ell+1}, ..., s_{t}$, observed by node $i$, learn a model at $i$ that predicts the subsequent $T$ states of $j$, $s_{t+1}, ..., s_{t+T}$.
\end{quote}
If the prediction of the power level received in transmissions from $j$ drops below the threshold for interference detection, $i$ can infer that $j$ will have moved out of its interference regions. Conversely, if a power level above the threshold is observed afresh, $i$ can infer that $j$ has entered its interference region. These observations can be used to expedite update of the channel hopping sequence (and also the schedule of $i$).

\subsection{Channel Hopping Sequence Assignment}


Link scheduling can be modeled as a vertex coloring problem on a conflict graph $\mathcal{G} = (\mathcal{V}, \mathcal{E})$, where $\mathcal{V}$ represents active communication links and $\mathcal{E}$ represents interference relationships. 
In standard formulations, finding the chromatic number of a general graph is NP-hard. However, our architecture operates under the condition that the total number of available channels $|U|$ is greater than or equal to $K$ (i.e., $|U| \ge K$), where $K$ is the number of active nodes transmitting traffic within an interference region. Consequently, the maximum degree of any vertex in the conflict graph, denoted as $\Delta$, is strictly bounded by $\Delta \leq K - 1$. Because the number of available colors (channels $|U|$) always strictly exceeds the maximum degree of the conflict graph ($\Delta < |U|$), this problem avoids NP-hard complexity. It reduces to a special case that can be optimally and efficiently solved using a standard greedy graph coloring algorithm\cite{assadi2018sublinear}.

\textbf{Time Complexity.} For a single active link to autonomously assign its local channel hopping sequence, the node must evaluate the predicted channel occupancy state across the sequence period $L$. In the worst-case scenario, the node scans all $|U|$ channels to find an unoccupied state for each of the $L$ time slots. Thus, the local scheduling time complexity for a single link is bounded by $\mathcal{O}(L \cdot |U|)$. For a network with $|E|$ active links, the greedy assignment resolves in linear time relative to graph size per time slot. Across period $L$, the global worst-case time complexity is $\mathcal{O}(L \cdot |E| \cdot |U|)$.

\section{M\lowercase{i}LAAP Solution Framework}\label{solution}
\vspace{-2mm}
As depicted in Fig.~\ref{MILAAP}, the \MILAAP framework consists of two {S}equence-to-{S}equence {A}ttention {N}etworks, $SSAN_{1}$ and $SSAN_{2}$, that solve the channel occupancy prediction and received power level prediction problems in Section~\ref{sec:PCO} respectively. $SSAN_{1}$ uses a single self-attention layer to capture the common period (or the least common multiple period) of the channel hopping sequences of nodes in the interference region to predict future channel occupancy vectors; $SSAN_{2}$ adopts a single multi-head self-attention layer to predict future received power levels from the nodes in the sensing region. 

\subsection{First-Phase Prediction of $CO$ Vectors and Capture of $CHS$ Period}

Given the assumption that all nodes in the interference region of node $i$ in a static network have channel hopping sequences of a common period $L$, if all of them transmit continually, it follows that $CO_{t}(i)$ is identical to $CO_{t+kL}(i)$, $k \in \mathbb{N}$. The functionality of $SSAN_{1}$, which includes capturing $L$ and using $L$ to predict the subsequent $CO$ vectors, is specified as follows.
The input to $SSAN_{1}$ is $X \in \left[0,1\right]^{\ell \times |U|}$, consisting of $\ell$ historical vectors of channel occupancy states observed by node $i$, $CO_{t-\ell+1}(i), ..., CO_{t}(i)$. $SSAN_{1}$ computes the queries ($M_{Q}$), keys ($M_{K}$), and values ($M_{V}$) on $X$ as:
\begin{equation}
\begin{array}{l}
M_{Q} := XW_{Q} \\
M_{K} := XW_{K} \\
M_{V} := XW_{V},
\label{SSAN_1_kqv}
\end{array}
\end{equation}
where $W_{Q}$, $W_{K}$, $W_{V} \in \mathbb{R}^{|U| \times |U|}$ are the learned linear mappings for the queries, keys, and values, respectively. The mappings are performed by the feedforward neural networks.

\begin{figure}[hbt!]
    \centering
     \subfigure
     {
        \includegraphics[width=0.6\textwidth]{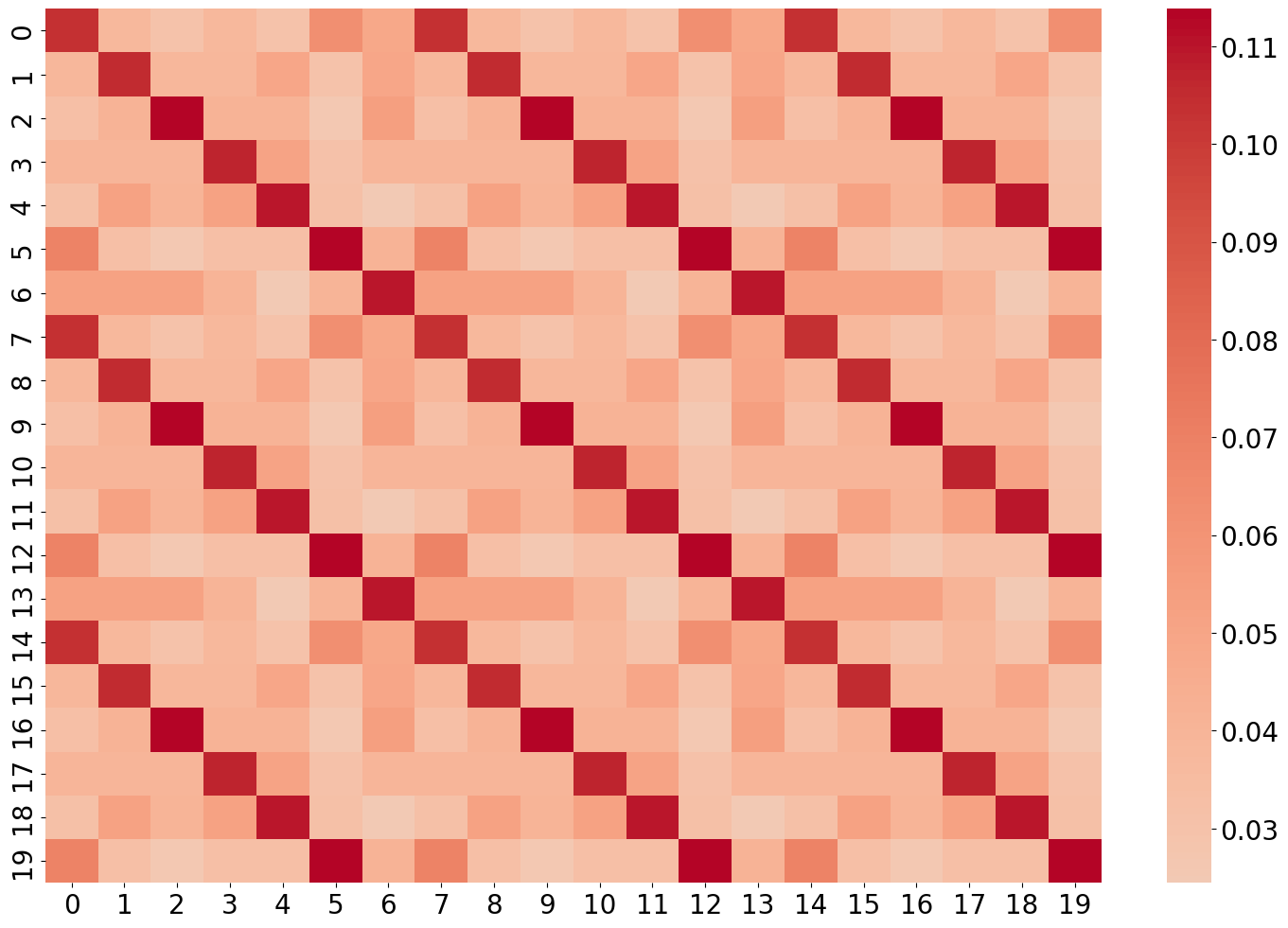}
    }
    \vspace{-4mm}
    \caption{Example of attention weights for the channel occupancy vectors of length 20 and $L=7$. There exists a pattern of period of 7 which can be used to predict the future channel occupancy vectors.}
    \label{attention_period_7}
\end{figure}

The calculation scheme of $SSAN_{1}$ is shown as follows:
\begin{equation}
\begin{array}{l}
scores := {\bf AttentionScore}(M_{Q} , M_{k}) \\
weights := {\bf Softmax}(scores) \\
period := {\bf CalculatePeriod}(weights) \\ 
M_{shi\!f\!ted} := {\bf MatrixShift}(M_{V} , period) \\
Y := {\bf BinOutput}(M_{shi\!f\!ted}W_{O}),
\label{SSAN_1}
\end{array}
\end{equation}
where $W_{O} \in \mathbb{R}^{|U| \times |U|}$ is the learnable weight matrix for the output $Y \in [0, 1]^{T \times |U|}$.
The attention scores are calculated first and {\bf Softmax} is used to normalize the scores to get a probability distribution (i.e., the attention weights). Then, we add a function, {\bf CalculatePeriod}, that captures the common period of the input vectors based on the attention weights. Fig.~\ref{attention_period_7} shows an example of an attention weight matrix computed for an input of 20 vectors with $L=7$. The channel hopping pattern period of 7, where $CO_{t} = CO_{t+7}, t=0,1, ..., 12$, is evidently modeled in the attention weights. The function captures the common period by determining the high weights (dark red cells) and then calculating the period (i.e., the distance between the high weights in each row) that appears most often. If the input length $\ell$ is not divisible by $L$, an offset is added for the following prediction. 
{\bf MatrixShift} operates to rollover the rows in $M_{V}$ based on the offset ($\ell \ \% \ L$) so that $M_{shi\!f\!ted}$ can then be used for predicting the output $Y$, as follows: 
\vspace{-2mm}
\begin{equation}
\begin{array}{l}
o\!f\!f\!set := \ell \ \% \ L \\
M_{shi\!f\!ted}[i] := M_{V}[(i+o\!f\!f\!set)] \ \% \ L, \; \; i = 0..T-1.\\
\end{array}
\end{equation}
{\bf BinOutput} rounds the elements below $0.5$ in a matrix into 0s and those above $0.5$ into 1s.  

\vspace{-6mm}
\subsection{Second-Phase Prediction of Received Power Levels}
Let $RP_{t}(i) \in \mathbb{R}^{|U|}$ denote the vector of received power level of all channels observed by node $i$ at time slot $t$. In addition to the $CO_{t-\ell+1}(i), ..., CO_{t}(i)$ vectors monitored from $t-\ell+1$ to $t$, node $i$ also collects $RP_{t-\ell+1}(i), ..., RP_{t}(i)$. 
Since the ID of node $j$ transmitting over a monitored channel $ch$ at time $t$ can be unknown to node $i$, it suffices that $i$  know that $\{RP_{t-kL}(i, ch) | k = 0 ... \lfloor{\frac{\ell}{L}}\rfloor-1 \}$ can be associated with the same node using the $L$ calculated in the first-phase prediction.\footnote{For the case where two nodes $j_1$ and $j_2$ use the same channel for communication at $t$, we assume the one with less distance to node $i$ dominates the received power level monitored at $i$.} Therefore, $\{RP_{t-kL}(i, ch) | k = 0 ... \lfloor{\frac{\ell}{L}}\rfloor-1 \}$ can be used as the historical data sequence associated with some nodes to predict its future received power levels.
$SSAN_{2}$ adopts a multi-head self-attention layer, in which each independent head has the opportunity to learn a different mobility pattern (e.g., a node coming into or vacating $i$'s interference region). The input to $SSAN_{2}$, $X$, is $\left[RP_{t-(\lfloor{\frac{\ell}{L}}\rfloor-1)L}(i, ch), RP_{t-(\lfloor{\frac{\ell}{L}}\rfloor-2)L}(i, ch), ..., RP_{t}(i, ch)\right] \in \mathbb{R}^{\lfloor{\frac{\ell}{L}}\rfloor}$.
$SSAN_{2}$ computes the queries ($M_{h}^{Q}$), keys ($M_{h}^{K}$), and values ($M_{h}^{V}$) for head $h$ according to Eq.~\ref{SSAN_1_kqv}. The calculation scheme in $SSAN_{2}$ is shown as follows:

\begin{equation}
\begin{array}{l}
scores_{h} := {\bf AttentionScore}(M_{h}^{Q} , M_{h}^{k}) \\
M_{h}^{W} := {\bf Softmax}(scores_{h}) \\
M_{O}:= {\bf Concat}(M_{1}^{W}M_{1}^{V}, ..., M_{n}^{W}M_{n}^{V}) \\
Y := {\bf BinOutput}(M_{O}W_{O}).
\label{SSAN_2}
\end{array}
\end{equation}
Each head separately calculates the attention scores and {\bf Softmax} is used to normalize the scores to get the attention weight matrix $M_{h}^{W}$. The results of all the heads are concatenated into a matrix $M_{O}$ and then output as the bit matrix $Y$, denoting the prediction result.

\textbf{Correction of $CO$ states.} The output $RP_{t+L}(i, ch)$ is used to correct the channel occupancy state $CO_{t+L}(i, ch)$ predicted in the first-phase. Let $\theta_{i}$ be the minimum received power level at node $i$ from neighbor $j$ transmitting in $i$'s interference region. $RP_{t+L}(i, ch) < \theta_{i}$ and $CO_{t+L}(i, ch) = 1$ imply that even though a node has moved out of the interference region of $i$ channel $ch$ is erroneously considered to interfere at $i$. In that case, $CO_{t+L}(i, ch)$ is corrected to 0. Conversely, $RP_{t+L}(i, ch) \geq \theta_{i}$ and $CO_{t+L}(i, ch) = 0$ represent that a node has entered the interference region of $i$ but is incorporated in the first-phase prediction. $CO_{t+L}(i, ch)$ is therefore corrected to 1.

\section{Performance Evaluation}
\vspace{-2mm}
We evaluated \MILAAP across two distinct mobile network applications: terrestrial wireless meshes and satellite meshes. The experimental results demonstrate that correcting the channel occupancy map via node mobility detection improves overall prediction accuracy.

\subsection{Experimental Settings}
We configure and evaluate the two distinct network environments as follows:

\subsubsection{Omnidirectional Terrestrial Wireless Meshes.}
The terrestrial wireless mesh is modeled as a Unit Disk Graph (UDG)\cite{von2005robust}, where the radio antennas are assumed to be omnidirectional. We simulate networks with nodes distributed over a given region using a uniform random distribution placement model\footnote{Let $R_{T}$ denote the transmission radius and $\rho$ denote the network density, which is the average number of nodes per ${R_{T}}^{2}$ area. Let $N$ be the number of nodes. It follows that all nodes are distributed in a square area whose side is of length $\sqrt{\frac{N \times {R_{T}}^{2}}{\rho}}$.}. To simulate concurrent communication that affects the channel occupancy states for a node's interference region, all networks initiate 10 random point-to-point traffic flows in which all routes follow the shortest path between the source and destination. Each communication node independently chooses a random channel hopping sequence with period $L$. The received power is calculated using Friis transmission equation, shown in Eq.~\ref{received_power_eq}.  The detailed experiment parameters are shown in Table~\ref{SimulationParams_IT_UDG}.

To evaluate \MILAAP across different types of node mobility in the terrestrial networks, we simulated three mobility models:

\begin{itemize}
    \item \textbf{Fixed Mobility (FM):} Each node selects a fixed velocity in $[0, 10]$ m/s and direction in $[0, 2\pi]$.
    \item \textbf{Random Waypoint (RWP):} Each node initially chooses a random location in a random direction $a_k \in [0, 2\pi]$ in the simulation area as the destination $D_k$, and then it moves towards the destination with a random velocity in $[0, 10]$ m/s. The node then assigns a new random destination $D_{k+1}$ and a random velocity when it arrives at $D_k$. This is repeated until the end of the simulation.
    \item \textbf{Smooth Random Waypoint (SRWP)~\cite{bettstetter2001smooth}:} The behavior is similar to RWP, but the assignment of the next destination $D_{k+1}$ and velocity $v_{k+1}$ depends on $a_k$ and $v_k$. A ratio $\epsilon$ is added to further limit the minimum and maximum values in the random assignment. $D_{k+1}$ is chosen to be a location in direction $a_{k+1} \in [a_{k+1} - 2\pi\epsilon, a_{k+1} + 2\pi\epsilon]$ and velocity $v_{k+1} \in [\max(0, (1-\epsilon)v_{k+1}), \min(10, (1+\epsilon)v_{k+1})]$.
\end{itemize}

We considered regions both with and without boundaries. In the bounded scenario, mobile nodes stop when they reach any boundary location.

\subsubsection{Directional Satellite Meshes.}
For the satellite mesh configuration, the network strictly comprises GEO-to-LEO inter-satellite links. Unlike the terrestrial setup, the radio antennas in the satellite network are highly directional, relying heavily on the specific beam widths to determine received power and interference. Regarding the mobility pattern, the satellite mesh consists of multiple orbits (orbital planes). Each satellite within the same orbit follows a fixed moving pattern defined by orbital mechanics \cite{tarhouni2025free}, maintaining its trajectory and predictable velocity over the simulated time steps. Similar to the terrestrial setup, the received power in satellite meshes is also calculated using Friis transmission equation. The detailed experiment parameters for setting the satellite mesh and the self-attention models are shown in Appendix~\ref{detailed_experiment_satellite}.

\subsection{Evaluation Results}
\subsubsection{Omnidirectional Terrestrial Wireless Meshes.}
\begin{figure}[hbt!]
    \centering
     \subfigure
     {
        \includegraphics[width=0.9\textwidth]{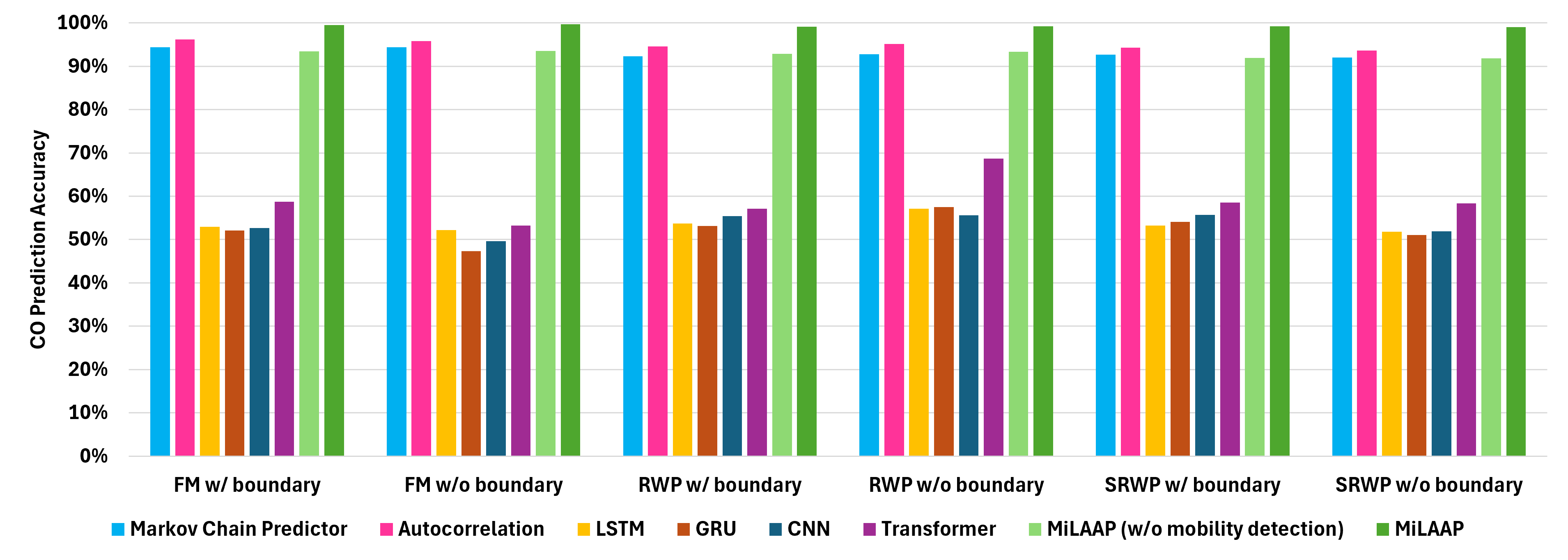}
    }
    \vspace{-4mm}
    \caption{Average channel occupancy ($CO$) prediction accuracy across various mobility patterns in terrestrial wireless meshes for non-ML, ML-based, and \MILAAP policies.}
    \label{comparison_UDG}
\end{figure}

For terrestrial wireless meshes, Figure~\ref{comparison_UDG} compares the performance of the ML-based models against non-ML historical solutions. The LSTM, GRU, and CNN models do not generalize well to the test dataset with unseen patterns in channel hopping sequences and node mobility, achieving an accuracy of only around 50\%. These models perform well on the training dataset primarily because they tend to simply memorize the seen patterns in the given CO sequences. Moreover, although the standard Transformer \cite{vaswani2017attention} possesses a higher capacity to learn complex patterns, it still leaves a large gap to optimality. This is because the standard Transformer lacks an additional component to independently handle the mobility predictions that affect the final observed CO states.

The historical predictors, Markov Chain Predictor and Autocorrelation, achieve above 90\% accuracy because they dynamically capture patterns based on recently observed CO states to predict future states. However, they exhibit a 4\% to 7\% gap to optimality because they lack the mechanism to detect the toggling of CO states on a channel caused by node mobility.

We present \MILAAP without mobility detection ($SSAN_{1}$ only) as an ablation study to evaluate the framework's performance without $SSAN_{2}$. This variant achieves results similar to those of non-ML historical predictors. By integrating $SSAN_{1}$ and $SSAN_{2}$, \MILAAP leverages $SSAN_{2}$ to capture node mobility patterns, effectively closing the performance gap to achieve ${\sim}100\%$ accuracy. It accomplishes this by correcting false positives (where neighbors have left the interference region) and false negatives (where neighbors have entered the interference region) in the CO predictions generated by $SSAN_{1}$.

{\bf Zero-shot generalizability and individual evaluations.}
Appendix~\ref{detailed_experiment_terrestrial} details the individual performance of $SSAN_{1}$ and $SSAN_{2}$. Both self-attention models achieve 100\% accuracy on their training datasets and exhibit near-optimal accuracy on unseen test datasets. Additionally, we investigate the zero-shot generalizability of \MILAAP across diverse channel hopping sequence periods and node mobility patterns, with the findings summarized in Table~\ref{attention1_results} and Table~\ref{milaap_generalizabiltiy}, respectively.

\subsubsection{Directional Satellite Meshes}
\begin{figure}[hbt!]
    \centering
     \subfigure
     {
        \includegraphics[width=0.9\textwidth]{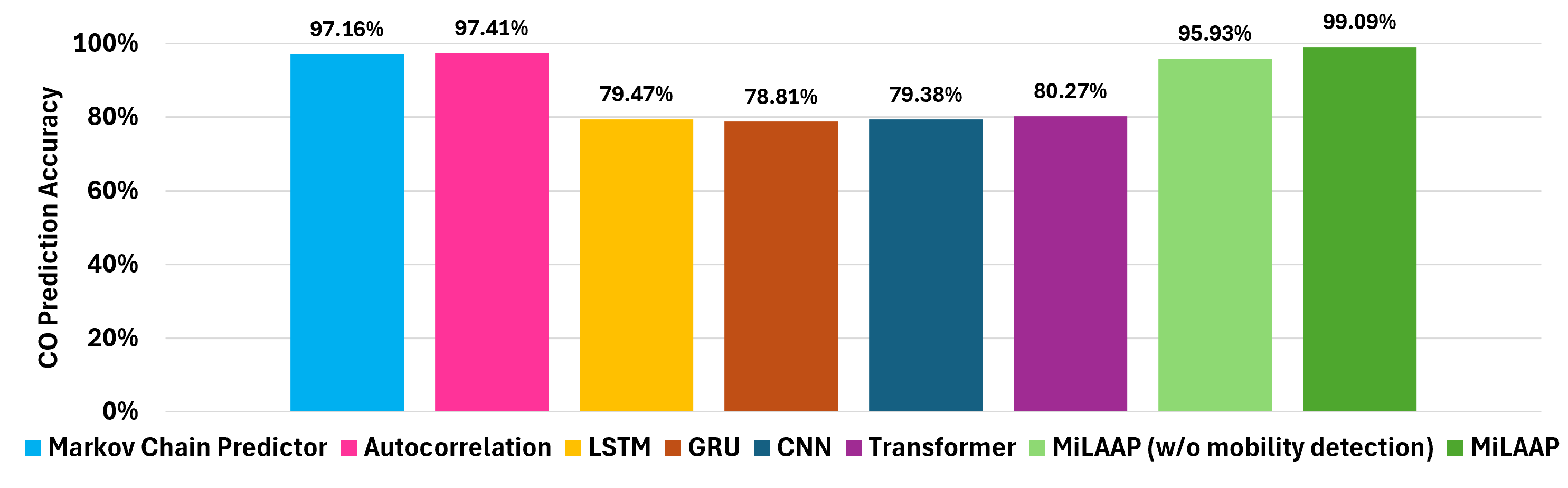}
    }
    \vspace{-4mm}
    \caption{Average channel occupancy ($CO$) prediction accuracy in satellite meshes for non-ML, ML-based, and \MILAAP policies.}
    \label{comparison_ISL}
\end{figure}

For satellite meshes, Figure~\ref{comparison_ISL} illustrates the prediction accuracy of \MILAAP and the baseline models under a more complex node mobility scenario. Unlike terrestrial wireless meshes (which are based on UDGs), where received power is dictated solely by the distance between the transmitter and the observer, the received power in satellite meshes depends on both the physical distance and the off-axis angle\footnote{This is the angle between the transmitting satellite's boresight (the vector from the transmitter to its target) and the outgoing interference vector (from the transmitter to the observer). This angle changes continuously over time as the transmitter, target, and observer are all in motion.}. 

The baseline models demonstrate performance trends similar to those observed in terrestrial wireless meshes. In \MILAAP, integrating $SSAN_{2}$ improves the accuracy by approximately 3\%, achieving a prediction accuracy of ${\sim}100\%$.

\section{Discussion}

{\bf Application to other network configurations.}
While our exposition associates channel hopping sequences with transmitters, \MILAAP can readily adapt to multi-channel multi-radio (MCMR) and multiple-input multiple-output (MIMO) communications. This flexibility supports diverse MAC and cross-layer objectives. Consequently, we deliberately omitted specific temporal scheduling, coordinated channel hopping sequence updates, and machine learning for temporal scheduling from Phase 3 (Figure \ref{MILAAP}).

{\bf Generalizable scheduling and comparable time complexity to historical predictors.}
\MILAAP offers zero-shot generalizability across various channel hopping sequence periods and node mobility patterns. Achieving $\sim \! \! 100\%$ channel occupancy prediction accuracy requires no additional retraining or fine-tuning of $SSAN_{1}$ and $SSAN_{2}$. Furthermore, this is accomplished without introducing additional time complexity compared to traditional solutions (e.g., historical predictors), and its calculations can be highly parallelized using GPUs, reducing the computational complexity to a logarithmic scale. A detailed complexity analysis for channel occupancy prediction across different models is provided in Appendix~\ref{complexity_CO_prediction}.

\section{Conclusion}
\MILAAP offers a lightweight machine learning framework for maintaining throughput efficiency in wireless networks subject to dynamic interference and mobility---such as in wireless IIoT and satellite networks---by locally updating channel hopping sequences via passive observation alone. Our validation demonstrates that training \MILAAP on a small dataset (from a single network with a fixed sequence period $L$) converges quickly, achieves high accuracy, and generalizes to a diverse set of channel hopping periods and node mobility patterns. 

\MILAAP maintains high prediction accuracy across diverse mobility models, particularly those exhibiting structured movement patterns; however, we expect prediction performance to degrade as node movements become increasingly random or highly frequent. The framework also demonstrates high accuracy for relatively long-lived traffic flows. Adapting \MILAAP for short-lived traffic or networks with highly intermittent node activity remains a topic for future research. Finally, while we did not explicitly validate the framework against external interference dynamics, the inherent self-stabilization of \MILAAP enables it to adapt to external interference, effectively avoiding channel collisions when externally induced channel changes persist or exhibit regular patterns.

%
%
%
\newpage
\bibliographystyle{splncs04}
\bibliography{mybibliography}

\newpage
\appendix
\renewcommand{\theHsection}{appendix.\Alph{section}}

\section{Notations}
Table~\ref{notation} summarizes the key notations used for channel occupancy prediction, received power level prediction, and the \MILAAP solution framework.

\begin{table}[ht]
\caption{Notations for \MILAAP}
\begin{center}
\begin{tabular}{c|c}
\hline
\textbf{Symbol}&\textbf{Meaning} \\
\hline
$U$ & set of all channels\\
\hline
$L$ & period of a channel hopping sequence\\
\hline
$i$ & node id\\
\hline
$(i, j)$ & link id\\
\hline
$ch\in U$ & channel id\\
\hline
$t$ & time slot number\\
\hline
$\Phi(i)$ & set of nodes in $i$'s interference region  \\
\hline
$CHS_{i}$ & channel hopping sequence for node $i$\\
\hline
$R_{T}$ & radius of reliable transmission region \\
\hline
$R_{I}$ & radius of interference region \\
\hline
$R_{S}$ & radius of sensing region\\
\hline
$CO_{t}(i) \in [0,1]^{|U|}$ & vector of monitored channel occupancy  \vspace*{-.5mm}  \\
 & states in $i$'s interference region at time $t$ \\
\hline
$CO_{t}(i, ch) \in [0,1]$ & $ch$'s channel occupancy state \vspace*{-.5mm}  \\
 & in $i$'s interference region at time $t$ \\
\hline
$RP_{t}(i) \in \mathbb{R}^{|U|}$ & vector of received power level \vspace*{-1mm} \\
 &  over channels observed at $i$ at time $t$ \\
\hline
$RP_{t}(i, ch) \in \mathbb{R}$ & $ch$'s received power level at $i$  at time $t$\\
\hline
$\theta_{i}$ & minimum received power level at $i$ from \vspace*{-.5mm} \\
& $j$ transmitting in $i$'s interference region \\
\hline
\end{tabular}
\label{notation}
\end{center}
\end{table}

\section{Detailed Experimental Result for Terrestrial Wireless Networks}
\label{detailed_experiment_terrestrial}
\subsection{Experiment Parameters}
The experiment parameters for setting the omnidirectional terrestrial wireless meshes and the self-attention models ($SSAN_{1}$ and $SSAN_{2}$) are listed in Table~\ref{SimulationParams_IT_UDG}.

\begin{table}[htbp]
\vspace*{4mm}
\caption{Experiment Parameters for $SSAN_{1}$ and $SSAN_{2}$ in Terrestrial Wireless Meshes}
\begin{center}
\begin{tabular}{c c c}
\hline
\textbf{Symbol}& \textbf{Meaning} & \textbf{Value} \\
\hline
$N$ & size of training and testing graph & 200 \\
$\rho$ & density of graph & 4 \\ 
$R_{T}$ & transmission radius & 1000m \\
$R_{I}$ & interference radius & 1000m \\
$R_{S}$ & sensing radius & 1100m \\
$\epsilon$ & smooth ratio in SRWP & 0.1 \\
$|U|$ & number of channels & 8 \\
$L$ & period of channel hopping sequences & 4 \\
$\ell$ & number of historical vectors as the input to \MILAAP & 40 \\
$\delta$ & sampling interval of $CO$ and $RP$ data & 1 sec \\
$\theta_{i}$ & received power threshold for interference detection & -80 dB \\
$X_{1}$ & shape of $SSAN_{1}$ input & $[0,1]^{40 \times 8}$ \\
$Y_{1}$ & shape of $SSAN_{1}$ output & $[0,1]^{40 \times 8}$ \\
$EpNum_{1}$ & \# of training epochs for $SSAN_{1}$ & 100 \\ 
$X_{2}$ & shape of $SSAN_{2}$ input & $\mathbb{R}^{10}$ \\
$Y_{2}$ & shape of $SSAN_{2}$ output & $\mathbb{R}$ \\
$H$ \ & number of heads used in $SSAN_{2}$ & 2 \\
$EpNum_{2}$ & \# of training epochs for $SSAN_{2}$ & 500 \\
\hline
\end{tabular}
\label{SimulationParams_IT_UDG}
\end{center}
\end{table}

\subsection{Testing First-Phase ($SSAN_{1}$) Prediction}
\label{test_SSAN_1}

\begin{table*}[!htbp]
\centering
\caption{The prediction accuracy for $SSAN_{1}$ versus other sequence-to-sequence models}
\begin{tabular}{*5c}
\hline
\multicolumn{1}{c}{Model} &  \multicolumn{1}{c}{Seen Periods} & \multicolumn{3}{c}{Prediction Accuracy} \\
\midrule
{} & {} & train dataset \ & \multicolumn{2}{c}{test dataset} \\
{} & {} & {} & seen periods \ & \ unseen periods \\\cline{3-5}

\multirow{2}{*}{$SSAN_{1}$} & 7 & 100\% & 99.51\% & 99.29\% \\\cline{2-5}
 & 5,7,9 & 100\% & 99.63\% & 99.31\% \\\hline
\multirow{2}{*}{Standard Self-Attention} & 7 & 97.50\% & 69.43\% & 67.01\% \\\cline{2-5}
 & 5,7,9 & 92.61\% & 70.79\% & 70.90\% \\\hline
\multirow{2}{*}{LSTM} & 7 & 100\% & 80.05\% & 62.51\% \\\cline{2-5}
 & 5,7,9 & 100\% & 78.17\% & 70.05\% \\
\hline
\end{tabular}
\label{attention1_results}
\end{table*}

We tested the ability of $SSAN_{1}$ to predict the future channel occupancy vectors at any given node.
We first generated training data samples for various periods $L$ of channel hopping sequences. (Within each data sample the channel hopping sequence associated with all nodes had a common $L$). 200 data samples were included in both the training and testing datasets.
We also generated two test datasets for testing generalizability: The first testing dataset includes seen periods of channel hopping sequences; i.e., these samples were not in the training dataset but their $L$ matched that of some samples in the training dataset. The second training dataset had only unseen periods of channel hopping sequences. The prediction accuracy we evaluated is defined as the average accuracy for predicting the future channel occupancy at the node.

Table~\ref{attention1_results} shows that $SSAN_{1}$ achieves a prediction accuracy close to 100\% in all classes of test samples. This validates the functionality of Eq.~\ref{SSAN_1} for deriving the period and calculating the period offset to predict future CO vectors. We also performed an ablation study with a standard self-attention that eschewed the functionality in Eq.~\ref{SSAN_1} to compare the ability to memorize seen periods. The decreased accuracy in the tests with unseen states demonstrates that the standard self-attention inadequately derives the period of channel hopping sequences to predict future CO vectors accurately. Also, a comparison test with an LSTM alternative test revealed that although the LSTM memorizes training data samples well (just as $SSAN_{1}$ does), the accuracy and generalizability of its prediction of future CO vectors in the test datasets is significantly worse than that of $SSAN_{1}$.

\begin{figure}[hbt!]
    \centering
    \subfigure[FM, Ground Truth]
    {
        \includegraphics[width=0.22\textwidth]{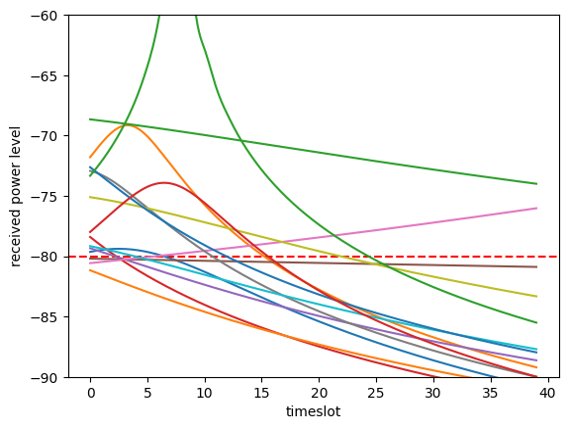}
        \label{FM_RP_ground_truth}
    }
    \subfigure[FM, $SSAN_{2}$ Prediction]
    {
        \includegraphics[width=0.22\textwidth]{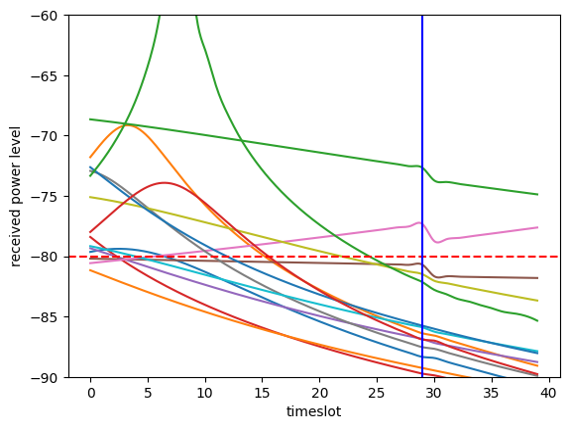}
        \label{FM_RP_prediction}
    }
    \subfigure[FM w/ boundary, Ground Truth]
    {
        \includegraphics[width=0.22\textwidth]{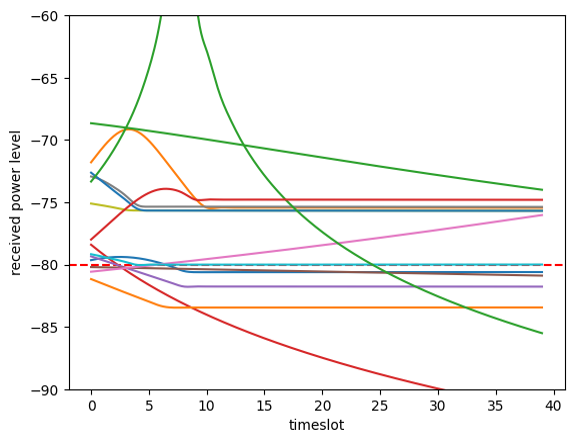}
        \label{FM_boundary_RP_ground_truth}
    }
    \subfigure[FM w/ boundary, $SSAN_{2}$ Prediction]
    {
        \includegraphics[width=0.22\textwidth]{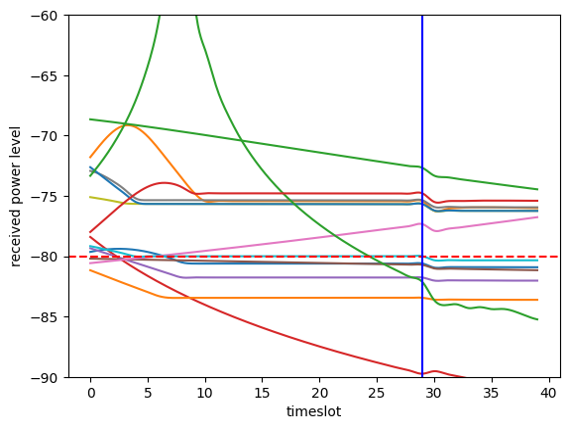}
        \label{FM_boundary_RP_prediction}
    }
    \\
    \subfigure[RWP, Ground Truth]
    {
        \includegraphics[width=0.22\textwidth]{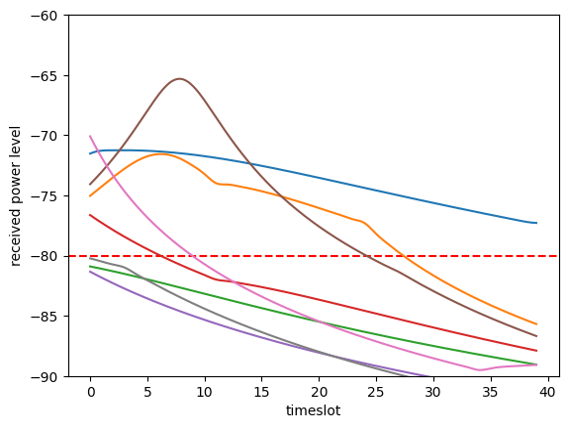}
        \label{RWP_RP_ground_truth}
    }
    \subfigure[RWP, $SSAN_{2}$ Prediction]
    {
        \includegraphics[width=0.22\textwidth]{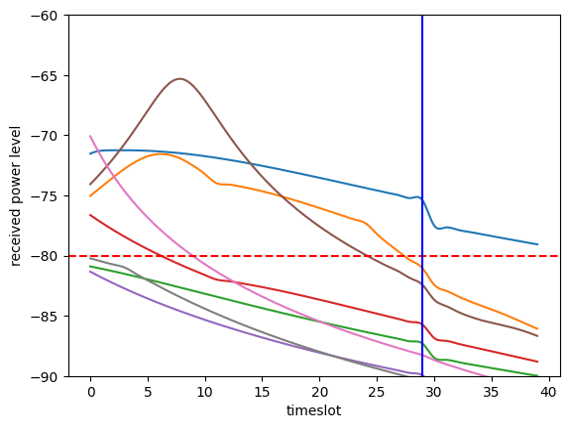}
        \label{RWP_RP_prediction}
    }
    \subfigure[RWP w/ boundary, Ground Truth]
    {
        \includegraphics[width=0.22\textwidth]{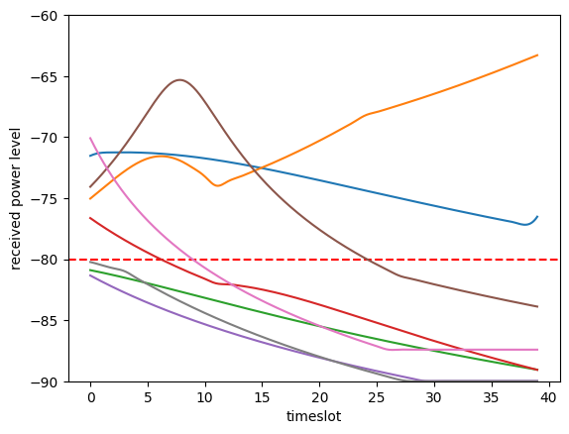}
        \label{RWP_boundary_RP_ground_truth}
    }
    \subfigure[RWP w/ boundary, $SSAN_{2}$ Prediction]
    {
        \includegraphics[width=0.22\textwidth]{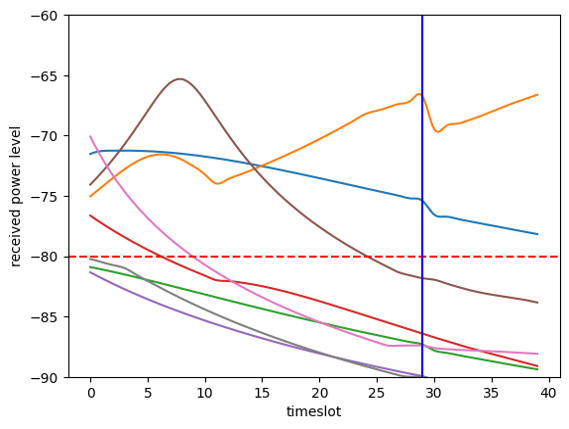}
        \label{RWP_boundary_RP_prediction}
    }
    \\
    \subfigure[SRWP, Ground Truth]
    {
        \includegraphics[width=0.22\textwidth]{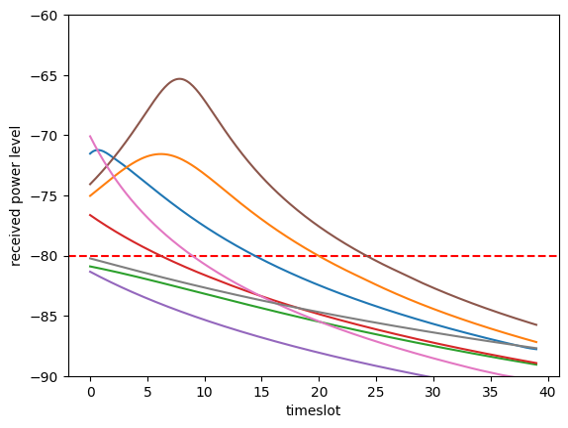}
        \label{SRWP_RP_ground_truth}
    }
    \subfigure[SRWP, $SSAN_{2}$ Prediction]
    {
        \includegraphics[width=0.22\textwidth]{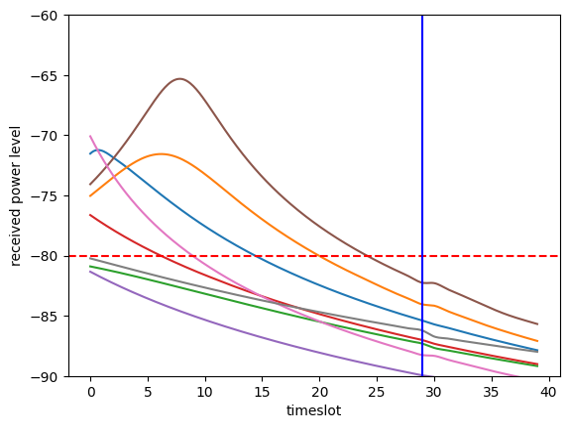}
        \label{SRWP_RP_prediction}
    }
    \subfigure[SRWP w/ boundary, Ground Truth]
    {
        \includegraphics[width=0.22\textwidth]{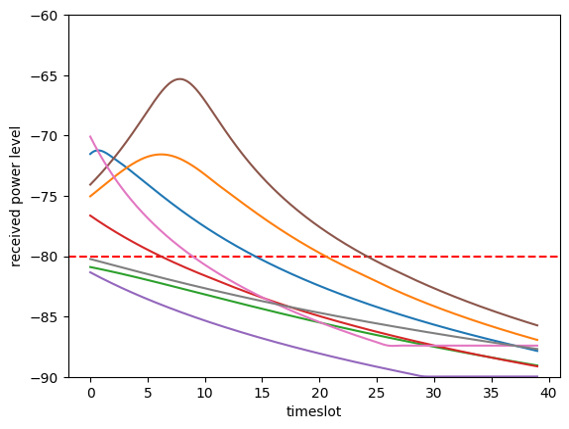}
        \label{SRWP_boundary_RP_ground_truth}
    }
    \subfigure[SRWP w/ boundary, $SSAN_{2}$ Prediction]
    {
        \includegraphics[width=0.22\textwidth]{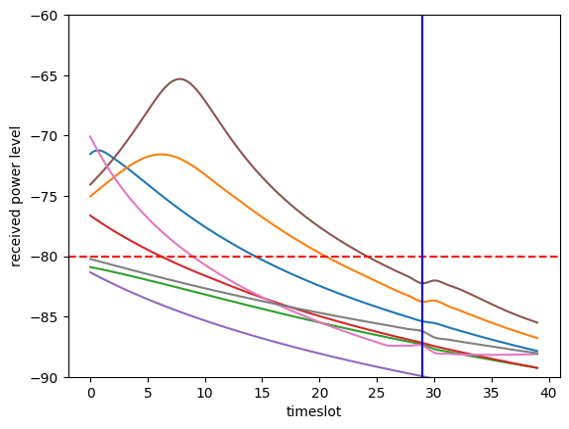}
        \label{SRWP_boundary_RP_prediction}
    }
    \caption{Examples of received power level at a node predicted by $SSAN_{2}$ for all mobile nodes in its sensing region, as compared with the ground truth for three mobility models. The blue solid vertical line denotes the time ($t=29$) from which the power level prediction starts. The red dashed horizontal line denotes the threshold $\theta_{i}$ (in dB) for interference detection.}
\label{RP_prediction}
\end{figure}

\subsection{Testing Second-Phase ($SSAN_{2}$) Prediction}
\label{test_SSAN_2}

\begin{table}[!htbp]
\centering
\vspace*{4mm}
\caption{The prediction accuracy for $SSAN_{2}$}
\begin{tabular}{*4c}
\hline
\multicolumn{2}{c}{Mobility Model} &  \multicolumn{2}{c}{Prediction Accuracy}\\
\midrule
{} & w/ boundary & train dataset & test dataset\\\cline{2-4}
\multirow{2}{*}{FM} & N & 100\% & 96.60\% \\\cline{2-4}
 & Y & 100\% & 96.09\% \\\hline
\multirow{2}{*}{RWP} & N & 100\% & 95.14\% \\\cline{2-4}
 & Y & 100\% & 96.40\% \\\hline
\multirow{2}{*}{SRWP} & N & 100\% & 98.63\% \\\cline{2-4}
 & Y & 100\% & 97.72\% \\
\hline
\end{tabular}
\label{attention2_results}
\end{table}

Next, we tested the prediction ability of $SSAN_{2}$ across the three different mobility models simulated in our dataset (FM, RWP, and SRWP mobility in networks with or without a boundary). For each network, we randomly choose a node $i$ to continuously monitor the received power level of all active nodes within its sensing range to obtain the input data for $SSAN_{2}$.
Recall that in the second phase, $i$ associates the received power levels in previous slots with different nodes $j$ in its sensing region. At each time slot $t$, $SSAN_{2}$ uses the measures of received power level from each node $j$ in the previous 30 slots, $s_{t-29}, ..., s_{t}$, to predict the received power level from $j$ in the next time slot, $s_{t+1}$, and to infer whether $j$ is within or outside the interference region by comparing that level with the threshold $\theta_{i}$. We define the prediction accuracy of $SSAN_{2}$ as the percentage that node $j$ is correctly predicted within (or outside of) $i$'s interference region.

Table~\ref{attention2_results} shows the prediction accuracy of $SSAN_{2}$ in networks with different mobility models. We further visualize the prediction results in Fig~\ref{RP_prediction}. In most cases, the predicted trajectories, starting from time slot 29 (after the solid blue lines), smoothly match the ground truth. We observe that mobile nodes in bounded networks yield a more complex mobility pattern compared to mobile nodes in unbounded networks. (The red and yellow flat line segments in Fig.~\ref{FM_boundary_RP_ground_truth} represent the case where nodes have stopped at the boundary; their received power level thus remains constant.) The added complexity however does not significantly affect the accuracy of the prediction. Together, these results imply that a multi-head self-attention mechanism suffices to capture the temporo-spatial dependencies needed to predict the trajectory of mobile nodes across a diverse set of mobility models. 

\subsection{Zero-shot Generalizability across Node Mobility Patterns}
We further study the zero-shot generalizability of \MILAAP, integrating $SSAN_{1}$ and $SSAN_{2}$, by evaluating how well a model trained on a single node mobility pattern adapts to unseen patterns. Note that we exclusively use bounded mobility scenarios for both the training and testing datasets, as unbounded scenarios represent simplified versions of these patterns. As shown in Table~\ref{milaap_generalizabiltiy}, even when trained on only one mobility pattern in terrestrial networks, \MILAAP achieves ${\sim}100\%$ channel occupancy prediction accuracy across all other mobility models. Specifically, training on FM, RWP, or SRWP yields a  performance of at least 98.83\% across all datasets. Notably, the \MILAAP model trained on the RWP (w/ boundary) dataset achieves the highest performance across all mobility patterns. This is because RWP represents the most complex mobility model and can be regarded as a superset of the FM and SRWP patterns.

\begin{table}[!htbp]
\centering
\vspace*{4mm}
\caption{The generalizability ($CO$ prediction accuracy) of \MILAAP under datasets with different mobility patterns}
\begin{tabular}{*4c}
\hline
\multicolumn{1}{c}{Train Dataset} & \multicolumn{3}{c}{Test Datasets}\\\hline
{} & FM w/ boundary \ \ & RWP w/ boundary \ \ & SRWP w/ boundary\\\cline{2-4}
FM w/ boundary & 99.44\% & 98.83\% & 99.08\% \\\hline
RWP w/ boundary & {\bf 99.57\%} & {\bf 99.12\%} & {\bf 99.36\%} \\\hline
SRWP w/ boundary & 99.36\% & 98.94\% & 99.06\% \\
\hline
\end{tabular}
\label{milaap_generalizabiltiy}
\end{table}

\section{Evaluation of Satellite Meshes}
\label{detailed_experiment_satellite}
The experiment parameters for the configuration of directional satellite network and the self-attention models ($SSAN_{1}$ and $SSAN_{2}$) are listed in Table~\ref{SimulationParams_Config_ISL} and Table~\ref{SimulationParams_IT_ISL}, respectively.

\begin{table}[htbp]
\vspace*{4mm}
\caption{Experiment Parameters for Satellite Network Configuration}
\begin{center}
\begin{tabular}{c c c}
\hline
\textbf{Symbol}& \textbf{Meaning} & \textbf{Value} \\
\hline
$\mathcal{O}_{GEO}$ & number of the GEO orbital planes & 1 \\
$\mathcal{N}_{GEO}$ & number of the GEO satellites per orbital plane & 4 \\
$\mathcal{M}_{GEO}$ & number of RF radios per GEO satellite & 4 \\
$\mathcal{T}_{GEO}$ & time period for GEO satellites to complete a circular orbit & 24 hours \\
$\mathcal{A}_{GEO}$ & altitude of the GEO satellites above the Earth & 35786 km \\
$\mathcal{O}_{LEO}$ & number of LEO orbital planes & 5 \\
$\mathcal{N}_{LEO}$ & number of LEO satellites per orbital plane & 5 \\
$\mathcal{M}_{LEO}$ & number of RF radios per LEO satellite & 1 \\
$\mathcal{T}_{LEO}$ & time period for LEO satellites to complete a circular orbit & 1.5 hours \\
$\mathcal{A}_{LEO}$ & base altitude for the first LEO plane in kilometers & 600 km \\
$\mathcal{A}^{+}_{LEO}$ & altitude increment added to each subsequent LEO plane & 5 km \\
$BM$ & beam width for the radio of GEO/LEO satellites & 30 degrees \\
$P_{TX}$ & transmission power the radio of GEO/LEO satellites & 20 dB \\
$G_{TX}$ & transmission antenna gain for the radio of GEO/LEO satellites & 25 dBi \\
$G_{RX}$ & receiving antenna gain for the radio of GEO/LEO satellites & 25 dBi \\
$F$ & operational frequency band & 2.4 GHz\\
\hline
\end{tabular}
\label{SimulationParams_Config_ISL}
\end{center}
\end{table}

\begin{table}[htbp]
\vspace*{4mm}
\caption{Experiment Parameters for $SSAN_{1}$ and $SSAN_{2}$ in Satellite Meshes}
\begin{center}
\begin{tabular}{c c c}
\hline
\textbf{Symbol}& \textbf{Meaning} & \textbf{Value} \\
\hline
$|U|$ & number of channels & 8 \\
$L$ & period of channel hopping sequences & 4 \\
$\ell$ & number of historical vectors as the input to \MILAAP & 40 \\
$\theta_{i}$ & received power threshold for interference detection & -150 dB \\
$\delta$ & sampling interval of $CO$ and $RP$ data & 3 sec \\
$X_{1}$ & shape of $SSAN_{1}$ input & $[0,1]^{40 \times 8}$ \\
$Y_{1}$ & shape of $SSAN_{1}$ output & $[0,1]^{40 \times 8}$ \\
$EpNum_{1}$ & \# of training epochs for $SSAN_{1}$ & 100 \\ 
$X_{2}$ & shape of $SSAN_{2}$ input & $\mathbb{R}^{10}$ \\
$Y_{2}$ & shape of $SSAN_{2}$ output & $\mathbb{R}$ \\
$H$ \ & number of heads used in $SSAN_{2}$ & 2 \\
$EpNum_{2}$ & \# of training epochs for $SSAN_{2}$ & 500 \\
\hline
\end{tabular}
\label{SimulationParams_IT_ISL}
\end{center}
\end{table}


{\bf Received Power Calculation.}
Accurately determining the received power is critical for predicting mobility and establishing interference bounds within the satellite mesh. In this environment, the calculation differs significantly from terrestrial networks due to the use of highly directional radio antennas, which means that received power and interference rely heavily on specific beam widths. Unlike terrestrial meshes where distance alone dictates power, the received power in satellite meshes depends on both the physical distance and the off-axis angle between the satellites.  The baseline distance-driven power loss is calculated using the Friis transmission equation, which factors in the transmission power, transmission and reception antenna gains, and the operational frequency over the 3D distance. Evaluated on a logarithmic scale (dB), the received power $RP$ is calculated as:  

\begin{equation}
\begin{array}{l}
RP = P_{TX} + G_{TX} + G_{RX} - PL.
\label{received_power_eq}
\end{array}
\end{equation}

In this formulation, $P_{TX}$ is the transmission power, $G_{TX}$ and $G_{RX}$ are the transmission and reception antenna gains, and $PL$ is the free space path loss. To account for the off-axis angle, the effective transmission gain is scaled down exponentially based on the angle between the intended target and the victim satellite, divided by the standard deviation of the beam width. Additionally, when evaluating interference, if two transmitting nodes operate on the same communication channel concurrently, the transmitter with less distance to the observing node dominates the received power level. 


\section{Complexity Analysis for Channel Occupancy Prediction}
\label{complexity_CO_prediction}
For the following complexity analyses, we define $|U|$ as the number of channels, $\mathcal{N}$ as the number of input $CO$ vectors (where each vector contains $|U|$ elements), and $\mathcal{M}$ as the length of the output $CO$ vectors. The time complexity for predictive models can be decomposed into training (or model construction) and inference phases.

\subsection{Historical Predictors}
For non-ML historical predictors, training is not conducted offline; rather, it occurs dynamically for each inference cycle. Given a set of input vectors, these models must actively reconstruct the state space or search for the channel hopping period to capture the underlying pattern. Once identified, this pattern dictates the prediction, necessitating that the entire training and inference process be repeated whenever a new prediction round is performed.

\subsubsection{Markov Chain Predictor.}
The Markov Chain Predictor operates by treating each historically observed channel occupancy vector as a discrete state within a probability model. During its dynamic model construction (training) phase, the algorithm sequentially processes the input vectors to record the unique transitions between states, effectively building a transition matrix that maps the likelihood of one state following another. For inference, the predictor uses the most recently observed state to look up the most probable subsequent state in the matrix, stepping forward iteratively to generate the predicted future sequence.

\begin{itemize}
    \item \textbf{Training Complexity:} $\mathcal{O}(\mathcal{N} \times |U|)$. This process is executed for each new input.
    \item \textbf{Inference Complexity:} $\mathcal{O}(\mathcal{M} \times |U|)$.
    \item \textbf{Space Complexity:} $\mathcal{O}(I \times U)$, where $I$ represents the number of unique transitions observed in the input ($I \leq \mathcal{N}$).
\end{itemize}

\subsubsection{Autocorrelation (Pattern Repeater).}
The Autocorrelation algorithm predicts future channel occupancy by identifying repeating cyclic behaviors within the historical data. During its period estimation (training) phase, it performs an iterative, sequential search across a defined search space to calculate the correlation between past input vectors and determine the dominant hopping period. Once this specific period is captured, the algorithm uses it as a pattern repeater, directly copying and shifting the historical sequence forward in time to infer future channel occupancy states.

\begin{itemize}
    \item \textbf{Training (Period Estimation) Complexity:} $\mathcal{O}(S \times \mathcal{N} \times |U|)$, where $S$ is the search space size for the channel hopping sequence periods. Typically, $S = \mathcal{N}$.
    \item \textbf{Inference (Prediction) Complexity:} $\mathcal{O}(\mathcal{M} \times |U|)$.
\end{itemize}

\subsection{\MILAAP (Single-Head Self-Attention)}
The \MILAAP architecture predicts channel occupancy with the following decomposed time complexities. It is important to note that the training effort for \MILAAP is a one-time process. Once trained, the self-attention model can be applied directly to any node in the network, which means that no additional training effort is involved during the inference process. The inference (attention phase) complexity is calculated based on the complexity of matrix multiplication for the steps in Eq.~\ref{SSAN_1}.


\begin{itemize}
    \item \textbf{Inference (Attention Phase) Complexity:} $\mathcal{O}(\mathcal{N}^2|U| + \mathcal{N}|U|^2)$.
    \begin{itemize}
        \item \textit{Linear Projections} ($XW_Q, XW_K, XW_V$): $\mathcal{O}(\mathcal{N}|U|^2)$, where $X \in \mathbb{R}^{\mathcal{N} \times |U|}$ and $W_Q, W_K, W_V \in \mathbb{R}^{|U| \times |U|}$.
        \item \textit{Attention Scores} ($M_{Q} M_{K}^{\top}$): $\mathcal{O}(\mathcal{N}^2|U|)$, where $M_{Q} \in \mathbb{R}^{\mathcal{N} \times |U|}$ and $M_{K}^{\top} \in \mathbb{R}^{|U| \times \mathcal{N}}$.
        \item \textit{Weighted Output} ($A_{scores} \times M_{V}$): $\mathcal{O}(\mathcal{N}^2|U|)$, where $A_{scores} \in \mathbb{R}^{\mathcal{N} \times \mathcal{N}}$ and $M_{V} \in \mathbb{R}^{\mathcal{N} \times |U|}$.
    \end{itemize}
    \item \textbf{Output Phase:} $\mathcal{O}(\mathcal{M} \times U)$.
\end{itemize}

\subsection{Complexity Analysis for Running \MILAAP on GPU}
When executing \MILAAP on GPU hardware, the matrix multiplications take advantage of parallel processing capabilities.

\subsubsection{Resource-Constrained GPU ($C < \max(\mathcal{N}^2|U|, \mathcal{N}|U|^2)$.} Let $C$ denote the number of GPU cores. The overall complexity scales to $\mathcal{O}(\frac{\max(\mathcal{N}^2|U|, \mathcal{N}|U|^2)}{C})$.
\subsubsection{Massively Parallel GPU ($C > \max(\mathcal{N}^2|U|, \mathcal{N}|U|^2)$.} When there are sufficient cores (like modern AI GPUs with Tensor Cores) to assign one core to every multiplication operation, the dot product's summation step becomes the bottleneck. On a parallel architecture, summing numbers takes logarithmic time via a tree-reduction algorithm, reducing the complexity drastically to $\mathcal{O}(\log |U| + \log \mathcal{N})$.

\subsection{Comparison of Time Complexity across Predictive Models}
Table~\ref{summed_complexity} summarizes the inference time complexity across the predictive models shown above. Typically, the output sequence length $\mathcal{M}$ does not exceed the input sequence length $\mathcal{N}$, as future state predictions rely entirely on patterns captured within the historical data. Consequently, the $\mathcal{M}\times|U|$ term is asymptotically dominated by the $\mathcal{N}\times|U|$ term and can be safely omitted when defining the overall time complexity bounds.

\begin{table}[htbp]
\centering
\caption{Summed Inference Time Complexity Comparison for Predictive Models}
\label{summed_complexity}
\renewcommand{\arraystretch}{1.5}
\begin{tabular}{|l|c|}
\hline
\textbf{Model} & \textbf{Summed Time Complexity (Entire Inference Phase)} \\ \hline
Markov Chain Predictor & $\mathcal{O}(\mathcal{N}\times|U|)$ \\ \hline
Autocorrelation & $\mathcal{O}(S\times\mathcal{N}\times|U|)$ \\ \hline
\MILAAP (Standard CPU) & $\mathcal{O}(\mathcal{N}^{2}|U|+\mathcal{N}|U|^{2})$ \\ \hline
\MILAAP (Constrained GPU) & $\mathcal{O}\left(\frac{\max(\mathcal{N}^{2}|U|,\mathcal{N}|U|^{2})}{C}\right)$ \\ \hline
\MILAAP (Parallel GPU) & $\mathcal{O}(\log|U|+\log\mathcal{N})$ \\ \hline
\end{tabular}
\end{table}

\end{document}